\documentclass[twocolumn,showpacs,aps,prb]{revtex4}
\input{epsf}
\begin{document}
\title{Theory of one-dimensional double-barrier quantum pump \\
in two-frequency signal regime.}

\author{M.M. Mahmoodian}
\email{mahmood@isp.nsc.ru} \affiliation{Institute of Semiconductor
Physics, Siberian Division, Russian Academy of Sciences,
\\Novosibirsk, 630090 Russia}

\author{M.V. Entin}
\email{entin@isp.nsc.ru} \affiliation{Institute of Semiconductor
Physics, Siberian Division, Russian Academy of Sciences,
\\Novosibirsk, 630090 Russia}


\begin{abstract}
A one-dimensional system with two $\delta$-like barriers or wells
bi-chromaticaly oscillating at frequencies $\omega$ and $2\omega$
is considered. The alternating signal leads to the direct current
across the structure (even in a symmetric system). The properties
of this quantum pump are studied in a wide range of the system
parameters.
\end{abstract}

 \maketitle

The quantum pump is a device that generates the stationary current
under action of alternating voltage; it is a subject of numerous
recent publications (for example,
\cite{Switkes,Kim1,Thouless,Brouwer,Moskalets02,Moskalets03,Zhu,Kim,recent,recent1}.
The quantum pump is essentially analogous to various versions of
the photovoltaic effect studied in details from the beginning of
the eighties \cite{Bel,we,Bas,Iv,Entin-Shepel}. The difference is
that the photovoltaic effect is related to the emergence of a
direct current in a homogeneous macroscopic medium (the only
exception is the mesoscopic photovoltaic effect), while the pump
is a microscopic object. From the phenomenological point of view,
the emergence of a direct current in the pump is not surprising,
since any asymmetric microcontact can rectify ac voltage. However,
analysis of adiabatic transport in the quantum-mechanical object
leads to new phenomena, such as quantization of charge transport
\cite{Thouless}. Just this, analytically solvable, adiabatic
approach was utilized in the most of studies of quantum pumps
\cite{Brouwer,Moskalets02,Moskalets03}. In the recent papers we
have carried out the extensive study of the simplest model of the
one-dimensional quantum pump, containing two delta-like
harmonically oscillating barriers/wells. This model demonstrates
rich behavior which is ruled by a variety of system parameters.
The present paper deals with similar system to which alternating
bi-chromatic voltages are applied. The system can be exemplified
by a quantum wire with two narrow gates. The stationary bias
between the source and the drain is supposedly absent.

The system has a variety of regimes of the pump operation,
depending on the system parameters, e.g. frequency and amplitudes.
The effect is sensitive to the phase coherence of alternating
signals and can exist even in symmetric systems. The stationary
current is possible also in the case of different amplitudes of
alternating fields. We have studied the system both analytically
and numerically. The analytical approach is based on the
perturbational (with respect to amplitudes ($u_{ij}$) of a.c.
signal) consideration. The current contains independent
contributions caused by $u_{ij}$ and an interference term. The
elastic, absorption and emission channels participate in the
process. The case of strong alternating signal was studied
numerically.

We mostly concentrate on the case of symmetric system as more
interesting by its phase sensitivity.

\subsection*{Basic Equations}
The considered model is described by the one-dimensional
time-dependent potential:
\begin{eqnarray}\label{1}
U(x)=(u_{11}\sin\omega t+u_{12}\sin2\omega t)\delta(x+d)+\nonumber\\
(u_{21}\sin\omega t+u_{22}\sin2\omega t)\delta(x-d).
\end{eqnarray} where $t$ is the time, $2d$ is the distance between
$\delta$-barriers (wells); quantities $u_{ij}$ are measured in
units of $\hbar/md$ ($m$ is the electron mass); $p$, $E$, and
$\omega$ are the momentum, energy, and frequency measured in units
of $\hbar/d$, $\hbar^2/2md^2$, and $\hbar/2md^2$, respectively.

The solution to the Schr\"{o}dinger equation with the potential
(\ref{1}) is searched in the form
\begin{widetext}
\begin{eqnarray}\label{wf}
\psi=\sum\limits_n \exp\left[-i(E+n\omega)t\right]\left\{
\begin{array}{rl}\delta_{n,0}\exp\left(\frac{ip_nx}{d}\right)+r_n\exp\left(-\frac{ip_nx}{d}\right), &
x<-d,\\a_n\exp\left(\frac{ip_nx}{d}\right)+b_n\exp\left(-\frac{ip_nx}{d}\right), & -d<x<d, \\
t_n\exp\left(\frac{ip_nx}{d}\right), & x>d.
\end{array}\right.
\end{eqnarray}
\end{widetext}
Here, $p_n=\sqrt{p^2+n\omega}$ and $p=\sqrt{E}$. The wave function
(\ref{wf}) corresponds to the wave incident on the barrier from
the left. (In the final formulas, we mark the directions of
incident waves by the indices ''$\rightarrow$'' and
''$\leftarrow$''). The form of solution (\ref{wf}) corresponds to
absorption (for $n>0$) or emission ($n<0$)  of $n$ field quanta by
an electron after interaction with the vibrating barriers; $n=0$
relates to the elastic process. Quantities $t_n$ and $r_n$ give
the corresponding amplitudes of transmission (reflection). If the
value of $p_n$ becomes imaginary, the waves moving away from the
barriers should be treated as damped waves, so that
$\mbox{Im}p_n>0$.

The transmission amplitudes obey the equations:
$t_n=e^{-i(p+p_n)}T_n$,
\begin{widetext}
\begin{eqnarray}\label{sis+}
\left({{\begin{array}{*{20}c}
{u_{12}u_{22}g_{n-2}}\\
{u_{12}u_{21}g_{n-2}+u_{11}u_{22}g_{n-1}}\\
{-i\,u_{12}e^{-2ip_{n-2}}+u_{11}u_{21}g_{n-1}-i\,u_{22}e^{-2ip_n}}\\
-u_{12}u_{21}g_{n-2}-i\,u_{11}e^{-2ip_{n-1}}-i\,u_{21}e^{-2ip_n}-u_{11}u_{22}g_{n+1}\\
-u_{12}u_{22}\left({g_{n-2}+g_{n+2}}\right)-u_{11}u_{21}\left({g_{n-1}+g_{n+1}}\right)+2ip_ne^{-2ip_n}\\
-u_{12}u_{21}g_{n+2}+i\,u_{11}e^{-2ip_{n+1}}+i\,u_{21}e^{-2ip_n}-u_{11}u_{22}g_{n-1}\\
~~i\,u_{12}e^{-2ip_{n+2}}+u_{11}u_{21}g_{n+1}+i\,u_{22}e^{-2ip_n}\\
u_{12}u_{21}g_{n+2}+u_{11}u_{22}g_{n+1}\\
u_{12}u_{22}g_{n+2}\\
\end{array}}}\right)
\bullet\left({{\begin{array}{*{20}c}
{T_{n-4}^\to}\\
{T_{n-3}^\to}\\
{T_{n-2}^\to}\\
T_{n-1}^\to\\
T_n^\to~~\\
T_{n+1}^\to\\
T_{n+2}^\to\\
T_{n+3}^\to\\
T_{n+4}^\to\\
\end{array}}}\right)=2ipe^{-ip}\delta_{n,0},
\end{eqnarray}\end{widetext} and
\begin{widetext}
\begin{eqnarray}\label{sis-}
\left({{\begin{array}{*{20}c}
{u_{12}u_{22}g_{n-2}}\\
{u_{11}u_{22}g_{n-2}+u_{12}u_{21}g_{n-1}}\\
{-i\,u_{22}e^{-2ip_{n-2}}+u_{11}u_{21}g_{n-1}-i\,u_{12}e^{-2ip_n}}\\
-u_{11}u_{22}g_{n-2}-i\,u_{21}e^{-2ip_{n-1}}-i\,u_{11}e^{-2ip_n}-u_{12}u_{21}g_{n+1}\\
-u_{12}u_{22}\left({g_{n-2}+g_{n+2}}\right)-u_{11}u_{21}\left({g_{n-1}+g_{n+1}}\right)+2ip_ne^{-2ip_n}\\
-u_{11}u_{22}g_{n+2}+i\,u_{21}e^{-2ip_{n+1}}+i\,u_{11}e^{-2ip_n}-u_{12}u_{21}g_{n-1}\\
~~i\,u_{22}e^{-2ip_{n+2}}+u_{11}u_{21}g_{n+1}+i\,u_{12}e^{-2ip_n}\\
u_{11}u_{22}g_{n+2}+u_{12}u_{21}g_{n+1}\\
u_{12}u_{22}g_{n+2}\\
\end{array}}}\right)
\bullet\left({{\begin{array}{*{20}c}
{T_{n-4}^\leftarrow}\\
{T_{n-3}^\leftarrow}\\
{T_{n-2}^\leftarrow}\\
T_{n-1}^\leftarrow\\
T_n^\to~~\\
T_{n+1}^\leftarrow\\
T_{n+2}^\leftarrow\\
T_{n+3}^\leftarrow\\
T_{n+4}^\leftarrow\\
\end{array}}}\right)=2ipe^{-ip}\delta_{n,0}.
\end{eqnarray}\end{widetext} Here, $g_n=\sin2p_n/p_n$.

Provided that electrons from the right and left of the pump are in
equilibrium, and that they have identical chemical potentials
$\mu$, the stationary current is
\begin{equation}\label{cur}
J=\frac{e}{\pi\hbar}\int
dE\sum_n(|T_n^\rightarrow|^2-|T_n^\leftarrow|^2)f(E)\theta(E+n\omega),
\end{equation}
where $f(E)$ is the Fermi distribution function, and $\theta(x)$
is the Heaviside step function. The current is determined by the
transmission coefficients with real $p_n$ only.

At a low temperature, it is convenient to differentiate the
current with respect to the chemical potential:
\begin{equation}\label{G}
{\cal G}=e\frac{\partial}{\partial \mu}J=G_0
\sum_n\theta(\mu+n\omega)(|T_n^\rightarrow|^2-|T_n^\leftarrow|^2)_{p=p_F}.
\end{equation}
Here $G_0=e^2/\pi\hbar$ is the conductance quantum and $p_F$ is
the Fermi momentum. The resultant quantity ${\cal G}$ can be
treated as a two-terminal photoconductance (the conductance for
simultaneous change of chemical potentials of source and drain).

\subsection*{Perturbations theory}

In low-amplitude limit the stationary current (its derivative
${\cal G}$) is proportional to $${\cal
G}\propto\alpha_1u_{11}u_{21}+\alpha_2u_{12}u_{22}+\alpha_3u_{11}^2u_{22}
+\alpha_4u_{21}^2u_{12}.$$

Let's consider the case of symmetric system with
$u_{11}=-u_{21}=u, \;u_{12}=-u_{22}=v$. The systems symmetry leads
to the dependence ${\cal G}\propto u^2v$. This contribution arises
from corresponding terms in the transmission coefficients $T_0,
T_{\pm 1}, T_{\pm 2}$:
\begin{eqnarray}\label{Gpt}
T_0\propto 1+A_0u^2v,\nonumber\\T_{\pm 1}\propto A_{\pm 1}u+B_{\pm 1}uv,\\
T_{\pm 2}\propto A_{\pm 2}u^2+B_{\pm 2}v.\nonumber
\end{eqnarray}

The quantities $A_0$, $A_{\pm 1}$, $B_{\pm 1}$, $A_{\pm 2}$,
$B_{\pm 2}$ depend on $g_{\pm 1}$, $g_{\pm 2}$ which contains one-
and two-photon singularities. The calculations support this
dependence.

\subsection*{Numerical results}

\begin{figure}[ht] \leavevmode
\centering{\epsfysize=6cm\epsfbox{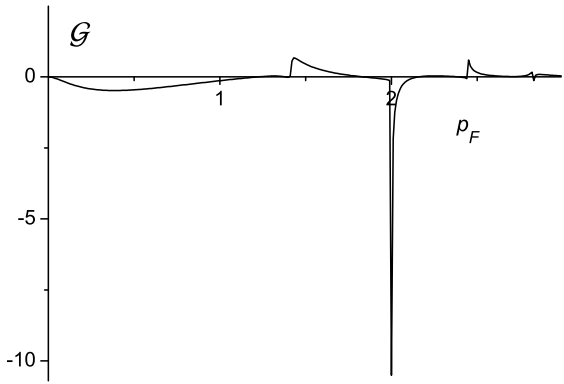}} \caption{The
derivative of the $\cal{G}$ {\it versus} the Fermi momentum. Here
$u_{11}=u_{12}=1,~u_{21}=u_{22}=-1,~\omega=2$. The left
delta-function oscillates from the barrier to zero hight, the
right delta-function corresponds to well whose depth changes from
zero value}\label{fig1}
\end{figure}

\begin{figure}[ht]
\leavevmode \centering{\epsfysize=6cm\epsfbox{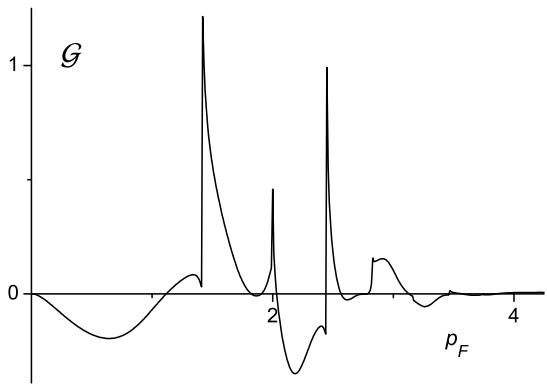}}
\caption{The dependence of $\cal{G}$ on the Fermi momentum. We set
$u_{11}=u_{12}=2,~u_{21}=u_{22}=-1,~\omega=2$.}\label{fig2}
\end{figure}

\begin{figure}[ht]
\leavevmode \centering{\epsfysize=6cm\epsfbox{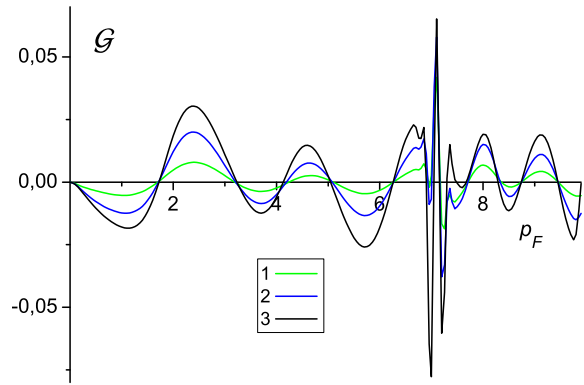}} \caption{
$\cal{G}$ {\it versus} the Fermi momentum;
$u_{11}=u_{12}=2,~u_{21}=u_{22}=-1,~\omega=50$ and different
$u_{11}=u_{12}$. The singularity at $p_F=7.1$ corresponds to the
single-photon threshold $p_F=\sqrt{\omega}$.}\label{fig5}
\end{figure}

\begin{figure}[ht]
\leavevmode \centering{\epsfysize=6cm\epsfbox{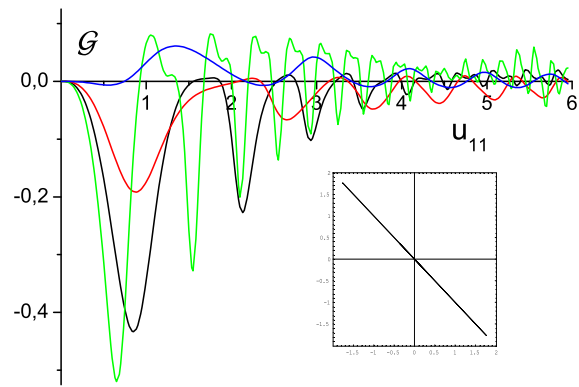}}
\caption{The dependence of $\cal{G}$ on the amplitudes
$u_{11}=u_{12}=-u_{21}=-u_{22}$ for $p_F=0.7$ and different signal
frequencies. For $u_{ij}\to 0$, ${\cal G}\propto u_{11}^3$.
Insert: trajectory in the phase space $\{u_1(t),u_2(t)\}$. The
enclosed area in the phase space is zero.}\label{fig6}
\end{figure}

\begin{figure}[ht]
\leavevmode \centering{\epsfysize=6cm\epsfbox{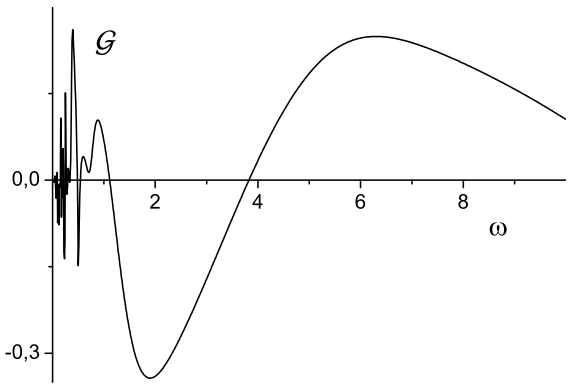}}
\caption{The dependence of $\cal{G}$ on the frequency for
$u_{11}=u_{12}=1$, $u_{21}=u_{22}=-1$, $p_F=0.7$.}\label{fig7}
\end{figure}

The figures \ref{fig1}-\ref{fig7} show typical plots of the
derivative of the stationary current with respect to the Fermi
energy $\partial J/\partial E_F={\cal G}\times 2e^2/h$ as a
function of the Fermi momentum (Figs. \ref{fig1}-\ref{fig5}), the
amplitude (Fig. \ref{fig6}) and the frequency (Fig. \ref{fig7}).

The figures~\ref{fig1}-\ref{fig5} present the quantity $\cal{G}$
as a function of the Fermi momentum for
$u_{21}=u_{22}=-1,~\omega=2$ and $u_{11}=u_{12}=1$ (Fig.
\ref{fig1}), 2 (Fig. \ref{fig2}). The peaks of the curves
correspond to the multi-photon threshold resonances with zero
energy state. The increasing of $u_{11}=u_{12}$ leads to the
strengthening of multi-photon singularities.

The curve in the Fig. \ref{fig1} corresponds to  antipodal signals
with same amplitudes $(u_{12}=-u_{22})$, that demonstrate the
importance of the phase coherency of the signals (in such
conditions the stationary current in the same system under
monochromatic voltage is vanishing \cite{recent}). The current
oscillates with the Fermi momentum due to the interference of the
electron waves in the structure. Besides, the current possesses
singularities caused by the resonances with the zero energy state
and their photon repetitions.

The figure \ref{fig5} shows the behavior of $\cal{G}$ for large
enough frequency where the only threshold singularity exists in
the concerned range of Fermi momentum values.

The figure \ref{fig6} depicts $\cal{G}$ {\it versus} $u_{11}$ for
the symmetric system ($u_{11}=u_{12}=-u_{21}=-u_{22}$). These
values are chosen so that there should be no current caused by
harmonic signals only (if $u_{11}=-u_{21}\neq 0, u_{12}=u_{22}=0$
or $u_{11}=u_{21}=0, u_{12}=-u_{22}\neq 0$).

Besides, the current vanishes in the adiabatic approximation which
is commonly used for consideration of quantum pumps. In the
adiabatic regime the charge transfer per a cycle is proportional
to the area covered by two parameters in the phase space. In our
case the parameters are $\{u_1(t),u_2(t)\}$. If
$u_{11}=u_{12}=-u_{21}=-u_{22}$ the trajectory in the phase space
$\{u_1(t),u_2(t)\}$ is a straight line and covers zero area (see
insert). Hence in the adiabatic ($\omega\to 0$) approximation
${\cal G}/\omega\to 0$.

At low $u_{ij}$ ${\cal G}\propto u_\omega^2u_{2\omega}$. This
behavior corresponds to coherent photovoltaic effect \cite{Bas1}.
At large $u_{ij}$ the amplitude dependence exhibits distinctive
oscillations which are determined by the commensurability of the
characteristic wavelength of excited electrons with the distance
between delta-functions. The curves for different frequencies
demonstrate splitting and beating of oscillations. Their amplitude
decays with $u_{ij}$. The oscillations period changes with the
frequency.

The figure \ref{fig7} demonstrates the dependence of $\cal{G}$ on
the frequency. The complicated structure of $\cal{G}$ in the low
frequency region is explained by multi-photon resonances reducing
for larger frequencies.

\section*{Conclusions}

The alternating voltage produces the stationary current by pumping
electrons between leads $x<-d$ and $x>d$. The effect is sensitive
to the phase coherence of alternating signals and can exist even
in symmetric systems. The stationary current is possible also in
the case of different amplitudes of alternating fields.

We have studied the system both analytically and numerically. The
analytical approach is based on the perturbational (with respect
to amplitudes of a.c. signal) consideration. The current contains
independent contributions caused by $u_{ij}$ and an interference
term. The elastic, absorption and emission channels participate in
the process. The case of strong alternating signal was studied
numerically.

Our calculations show that the stationary current is a
sophisticated function of the parameters, that reflects the
interference effects, presence of virtual states, and threshold
singularities.

The motivation to consider the symmetric case is that in this case
the stationary current caused by a single frequency is suppressed.
Only two intercoherent frequencies can cause the stationary
current. Hence, the system with specially symmetrized barriers can
be used for heterodyning of signals with $\omega$ and $2\omega$
frequencies and for frequency binding of similar laser sources.
The controllability of the system permits to realize the symmetric
conditions intentionally with good accuracy.

The work was supported by grants of the Program for support of
scientific schools of the Russian Federation No. 4500.2006.2, the
grant of the President of the Russian Federation No.
MK-8112.2006.2 and the Dynasty Foundation. We are grateful to
L.~S.~Braginsky for useful discussions.

\end{document}